\documentclass[floatfix, preprint, showpacs, showkeys, preprintnumbers, nofootinbib, superscriptaddress]{revtex4-1}
\usepackage{comment}
\usepackage{amssymb}
\usepackage{color}
\usepackage{rotating}
\usepackage{amsmath}
\usepackage{ulem}
\usepackage{overpic}
\usepackage{graphicx}
\usepackage{dcolumn}
\usepackage{bm}
\usepackage{chngpage}
\usepackage{multirow}
\usepackage{slashed}
\usepackage{indentfirst}
\usepackage{booktabs}
\usepackage{amssymb,amsfonts}
\usepackage{amsmath}
\usepackage{color}
\usepackage[colorlinks,citecolor=blue,linkcolor=blue,hypertex,breaklinks=true]{hyperref}

\begin{document}
\title{Input-driven analysis in predicting nuclear charge radii using Monte Carlo dropout Bayesian neural network}

\author{Jia-Xin Liu}
\affiliation{School of Physics, Ningxia University, Yinchuan 750021, China}

\author{Zhen-Yan Xian}
\affiliation{School of Physics, Ningxia University, Yinchuan 750021, China}

\author{Yan Ya}
\affiliation{School of Physics, Ningxia University, Yinchuan 750021, China}

\author{Si-Ying Ma}
\affiliation{School of Physics, Ningxia University, Yinchuan 750021, China}

\author{Na Tang}
\email[Corresponding author:]{natang@nxu.edu.cn}
\affiliation{School of Physics, Ningxia University, Yinchuan 750021, China}
\affiliation{Key Laboratory of Beam Technology of Ministry of Education, School of Physics and Astronomy, Beijing Normal University, Beijing 100875, China}

\author{Rong An}
\email[Corresponding author:]{rongan@nxu.edu.cn}
\affiliation{School of Physics, Ningxia University, Yinchuan 750021, China}
\affiliation{Key Laboratory of Beam Technology of Ministry of Education, School of Physics and Astronomy, Beijing Normal University, Beijing 100875, China}
\affiliation{Guangxi Key Laboratory of Nuclear Physics and Technology, Guangxi Normal University, Guilin, 541004, China}

\date{\today}

\begin{abstract}
 Nuclei charge radii play an essential role in understanding the fundamental interactions of finite quantum fermion systems. In this work, input-driven Bayesian neural network based on the Monte Carlo dropout approach has been built to characterize the systematic evolution of charge radii of nuclei with proton number $Z\geq20$ and mass number $A\geq40$.
  The motivated underlying mechanisms have been introduced into the input structures, which contain pairing effect, isospin asymmetry degree, the correlations between the valence nucleons and valence holes for neutron and proton, quadrupole deformation parameter $\beta_{20}$, and the local shape staggering phenomena of $^{181,183,185}$Hg isotopes.
  In addition, shell quenching effect is also taken into account by incorporating the modified Casten factor $P^{*}$ into the input structure.
  The quadrupole deformation parameters $\beta_{20}$ derived from finite-range droplet model (FRDM), relativistic mean field (RMF) theory and Weizs\"{a}cker-Skyrme (WS) approach are employed to analyze the local variations of nuclear charge radii.
  The hyperparameter is adjusted automatically in the constructed model.
  The calibrated results give comparable root-mean-square deviations (RMSD) in the training and validation sets with various shape deformation inputs.
  The abrupt increase in charge radii around $N = 60$ is well reproduced along $Z = 37$-$40$ isotopic chains, but this trend is less pronounced along $Z = 36$ and $41$ chains.
  This provides a indicator to confirm the rapid shape-phase transition regions around $N=60$ from the perspective of finite nuclei size.
  Shell quenching effect of charge radii along the bismuth isotopes are reproduced well at $N=126$, but slight deviations can be encountered due to the absence of high-order octupole deformation around $N=130$ regions and shape-staggering phenomena toward neutron-deficient regions, respectively.
  This means that more underlying mechanisms should be considered in the input layer.
\end{abstract}

%\pacs{25.70.Jj,24.10.-i}

\maketitle
\section{INTRODUCTION}\label{first}
%Reliable description of nuclear charge radii with high-precision is of fundamental importance in both nuclear physics and astrophysics.
Providing a reliable and high-precision description of nuclear charge radii is crucial for research in both nuclear physics and astrophysics.
Charge radii of finite nuclei show regular and irregular features along a long isotopic family.
The commonly observed characteristics associated with shell quenching phenomena can be generally found at the neutron numbers $N=28$, $50$, $82$, $126$, in which the shrinking trend appears close to the fully filled shells.
Meanwhile, the rapid increase is naturally presented across the corresponding neutron magic numbers $N$.
Between two adjacent neutron magic numbers, the inverted parabolic-like shape in charge radii can be shown obviously along a long isotopic family~\cite{ANGELI201369,LI2021101440}, but the amplitude is gradually weakened with the increasing proton numbers.
Furthermore, charge radii of neutron-rich isotopes beyond the neutron-closed shells appear to increase with almost similar trend~\cite{GarciaRuiz:2019cog,PhysRevC.105.L021303,sltw-g4d6}.
Another sign of the regular behavior in nuclear charge radii is the normal odd-even staggering (OES) phenomena.
Definitely, nuclear charge radii of even-$N$ isotopes are larger than the average of their corresponding odd-$N$ neighbors.

The discontinuous variations of charge radii are also generally presented in the specific regions throughout the whole nuclear chart.
From the nuclear mass perspective, the neutron number $N=20$ has been identified as being traditional magicity.
Actually, across $N=20$, the abrupt change of charge radii cannot be observed visibly in the Ar~\cite{KLEIN19961}, K~\cite{TOUCHARD1982169,PhysRevLett.79.375,PhysRevC.92.014305}, and Ca~\cite{NP-12-594} isotopes.
Unexpectedly, recent study suggests that the abruptly increasing trend can be found in charge radii of Sc isotopes across $N =20$~\cite{PhysRevLett.131.102501}. However, as encountered in K and Ca isotopes, the similar trend can be obviously displayed close to $N=28$ along the Sc isotopic chain due to the seniority symmetry limit~\cite{PhysRevLett.134.182501}.
In addition, the sudden increase of charge radii in specific regions, such as the onset of deformation at the neutron numbers $N = 60$ and $N = 90$, can be significantly observed due to shape-phase transition~\cite{LALAZISSIS199635,PhysRevLett.54.1991,PhysRevLett.117.172502}.
Furthermore, as is well known, more larger odd-even staggering in charge radii can be dramatically observed in the neutron-deficient isotopes around the proton number $Z = 82$ isotopic chain~\cite{ANSELMENT1986471,PhysRevC.95.044324,Marsh2018,PhysRevC.104.024328,PhysRevLett.126.032502}.
This attributes to the underlying mechanism of the abrupt shape-phase staggering phenomena.

With advanced experimental facility, laser spectroscopy method provides a potential access in extracting the charge radii of nuclei close to nucleons drip line~\cite{CAMPBELL2016127,YANG2023104005}.
The resultant researches suggest that much more charge radii data of nuclei far from $\beta$-stability line have been compiled~\cite{ANGELI201369,LI2021101440}.
Unfortunately, apparent discrepancy between the experimental detection and theoretical validation is encountered close to drip line region.
Actually, nuclear charge radii can be significantly influenced by various underlying mechanisms, such as pairing correlations~\cite{PhysRevC.104.064313,PhysRevLett.121.102501,PhysRevC.105.L021301}, shell quenching effect~\cite{KREIM201497,PhysRevLett.122.192502,PhysRevC.102.051303,BAGCHI2019251,PhysRevLett.129.142502,PhysRevC.110.034315}, shape deformation~\cite{LALAZISSIS199635,PhysRevLett.54.1991,PhysRevLett.117.172502,RevModPhys.83.1467,An_2023,PhysRevC.108.024310,1kbz-rx3m}, neutron-proton interaction~\cite{PhysRevC.107.064321}, intrinsic electromagnetic structure~\cite{PhysRevC.110.064319}, center-of-mass correction~\cite{PhysRevC.108.054314,PhysRevC.109.054323}, etc.
This means that available description of nuclear charge radii with high-precision should appropriately capture the motivated underlying mechanisms at much as possible.

To develop a unified method, nuclear model must not only reproduce known charge radii data but also be capable of effective extrapolation, thereby providing reliable guidance for future experimental detection.
Plenty of potential methods, which cover the empirical formulas~\cite{DUFLO199429,Zhang:2001nt,Piekarewicz2010,PhysRevC.87.024310,PhysRevC.87.054323,PhysRevC.88.011301,PhysRevC.89.024318,
Sheng2015,PhysRevC.94.064315}, local-relation based models~\cite{PhysRevC.95.014307,PhysRevC.102.014306,PhysRevC.104.014303}, chiral effective field theory~\cite{PhysRevLett.99.042501,PhysRevC.83.031301,PhysRevLett.110.242501,PhysRevC.91.051301,PhysRevC.101.014318}, relativistic energy density functional (EDF) theory~\cite{geng2003,PhysRevC.102.024307,PhysRevC.109.064302,b6j7-6q8z}, non-relativistic Skyrme EDF approach~\cite{PhysRevLett.102.242501,PhysRevC.82.035804,nyn5-69s3}, and the Fayans EDF model~\cite{FAYANS200049,PhysRevC.95.064328}, etc, have been built to describe the systematic trend of nuclear charge radii.
Actually, the extrapolated ability of the empirical formulas and local-relation based models is limited close to drip line due to the absence of data in the validation procedure.
Meanwhile, the chiral effective field theory and energy density functionals need more computing power or waste more computational time compared to the empirical formulas, although more underlying mechanisms can be captured from microscopic aspects.
In addition, it should be clearly mentioned that most of the developed theoretical models cannot give the quantified uncertainty.

Recently, machine learning approaches have achieved great success in nuclear physics and astrophysics.
Particularly, the global trend of changes of charge radii can be described well within various machine learning methods, which includes the
naive Bayesian probability classifier~\cite{PhysRevC.101.014304}, kernel ridge regression (KRR) model~\cite{Ma_2022},
artificial neural networks (ANNs)~\cite{Akkoyun_2013,PhysRevC.102.054323,PhysRevC.108.034315,PhysRevC.110.014308,PhysRevC.111.064304},
support vector regression (SVR) model~\cite{Jalili_2024}, convolutional neural networks (CNNs)~\cite{Cao2023,sym15051040}, and Bayesian neural networks (BNNs)~\cite{Utama_2016,PhysRevC.105.014308,DONG2023137726,PhysRevC.110.014316,vj25-zwd3,XIAN2025139662}, etc.
The Bayesian neural networks deduced from the combination of an artificial neural network and the Bayesian statistical theory can give the error uncertainty in the training and validated simulations.
This means that the constructed Bayesian neural network should capture the aleatoric and epistemic uncertainties naturally~\cite{Zhang10662945}.
Therefore, more motivated underlying mechanisms should be captured adequately in the calibrated protocol.
In Ref.~\cite{Utama_2016}, the BNN method simulates the residuals between the theoretical predictions and experimental charge radii data, and gives the root-mean-square deviation (RMSD) about $0.02$ fm.
The improved BNN approach, which incorporates the pairing and shell closure effects into the input structure, can reduce the RMSD to $0.015$ fm in simulating the charge radii data of nuclei with proton number $Z\geq20$ and mass number $A\geq40$~\cite{PhysRevC.105.014308}.
The further researches suggest that an available description of nuclear charge radii can be actually influenced by input structures~\cite{DONG2023137726,XIAN2025139662}.
All of these suggest that present machine learning algorithms with proper input-driven of physical features is desired toward a competitive accuracy in describing the systematic evolution of nuclear charge radii.

As mentioned earlier, the discontinuous behavior of nuclear charge radii can be used to characterize the spatial structural information, such as abrupt shape phase transition at $N=60$~\cite{PhysRevC.82.061302,XIANG20121,RODRIGUEZGUZMAN2010202,PhysRevC.85.034337,PhysRevC.85.034321,Kumar2021,PhysRevC.111.034306} and shape staggering phenomena in the neutron deficient regions around $Z=82$ isotopic chains~\cite{ANSELMENT1986471,PhysRevC.95.044324,Marsh2018,PhysRevC.104.024328,PhysRevLett.126.032502}.
This means that shape deformation effect plays an indispensable role in reproducing the local variations of charge radii of exotic nuclei~\cite{RevModPhys.83.1467}.
To facilitate the understanding of shape deformation on determining the systematic trend of changes of nuclear charge radii, the Monte Carlo dropout Bayesian neural network (MCD-BNN) models are constructed by varying input-driven quadrupole deformation contents.
In this work, input-driven analysis is performed by calibrating the charge radii of nuclei with proton number $Z\geq20$ and mass number $A\geq40$.
The motivated underlying mechanisms, which contains the fundamental building blocks with specific proton and neutron numbers, pairing effect, quadrupole deformation $\beta_{20}$, isospin asymmetry degree, local shape staggering effect of $^{181,183,185}$Hg, and the coupling terms between the valence particles and valence holes for neutron and proton, respectively, have been incorporated into the input structure.
In addition, the modified Casten factor $P^{*}$ is also used to depict shell quenching effect in nuclear charge radii.
To feature the influence of various deformation inputs on determining the local variations of charge radii, the detailed trend of changes of charge radii along the krypton, rubidium, strontium, yttrium, zirconium and bismuth isotopes are displayed in the corresponding discussion.

The structure of the paper is the following.
In Sec.~\ref{second}, the constructed MCD-BNN framework is presented succinctly.
In Sec.~\ref{third}, the numerical results about the RMSDs between the experimental data and the calibrated simulations, global trend of changes of charge radii along the krypton, rubidium, strontium, yttrium and zirconium isotopes are depicted. In addition, charge radii of the bismuth isotopes are also analyzed with various input quadrupole deformation parameters and the corresponding discussion has been provided.
Finally, a summary and outlook is given in Sec.~\ref{fourth}.

\section{THEORETICAL FRAMEWORK}\label{second}
The Monte Carlo dropout Bayesian neural network (MCD-BNN) model has achieved great success in analyzing various physical quantities with the quantified uncertainty, such as the mass distributions of the induced
fission~\cite{Huo2023}, spectral function~\cite{PhysRevResearch.4.043082},  and electron-carbon scattering data~\cite{PhysRevC.110.025501}, etc.
In contrast to traditional neural network, the over-fitting puzzle can be avoided by using the dropout method through the regularization technique~\cite{JMLRv15srivastava14a}.
Meanwhile, it should be noted that Monte Carlo dropout (MCD) approach plays an equivalent role in evaluating a neural network compared to the variational inference Bayesian learning method~\cite{Gal2016}.

In the MCD-BNN framework, the dropout architecture with a single hidden layer can be expressed as follows,
\begin{eqnarray}\label{bnn1}
\hat{y}=\sigma(xz_{1}W_{1}+b)z_{2}W_{2}.
\end{eqnarray}
Here, the quantity $\sigma$ denotes the nonlinear activation function, $x$ represents input data, $W_{1}$ stands for the weight matrix which links input layer and hidden layer, $b$ denotes the bias vectors, and $W_{2}$ represents the weight matrix connecting the hidden layer to output layer. The quantities $z_{1}$ and $z_{2}$ are binary vectors in the sampling process.
In the multiplying hidden activations, the Monte Carlo dropout approach is driven by the Bernoulli distributed random variables.
The random variables take the value of $1$ with probability parameter $p$ and $0$ for the invalidity of neuron in a given input~\cite{Gal2016,Gal20161}.

The regularization approach mentioned above has been used in the MCD-BNN model.
The $L_{2}$ regularization weighted by some weight decays provides a minimization objective shown as follows~\cite{Huo2023,Wen2020},
\begin{eqnarray}\label{bnn3}
L_{\mathrm{2}}=\frac{1}{N}\sum_{n=1}^{N}||y_{n}-\hat{y}_{n}||_{2}^{2}+\lambda_{\mathrm{decay}}\sum_{i=1}^{L}(||W_{i}||_{2}^{2}+||b_{i}||_{2}^{2}).\nonumber\\
\end{eqnarray}
The first term represents the Euclidean loss function used in the training and optimization process of neural network.
Here, the expression $\hat{y}_{n}$ denotes the neural network output with $L$ layers and $y_{n}$ shows the target value. The quantity $N$ is being the number of training data points. A regularization term is often added into the optimization process as shown in the second term.
Here, the expression $W_{i}$ denotes the weight matrixes and the bias vectors of dimensions for each layer ($i=1, 2, \cdots, L$) are represented by $b_{i}$.
The hyperparameter $\lambda_{\mathrm{decay}}$, which is automatically adjusted, serves as a regularization parameter to reduce the risk of over-fitting~\cite{Huo2023}. Note that the resultant hyperparameters $\lambda_{\mathrm{decay}}$ differ across the models constructed in this work.

\begin{figure}[htbp]
\includegraphics[scale=0.45]{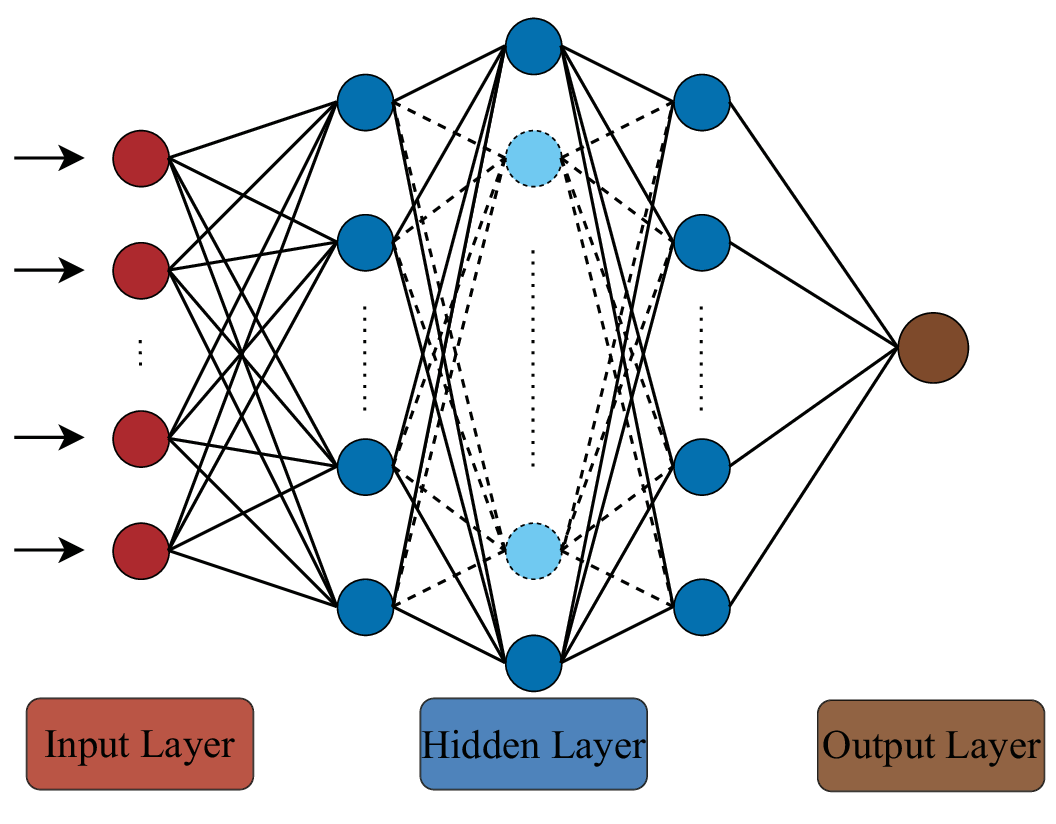}
\caption{Structure of the Monte Carlo dropout Bayesian neural network (MCD-BNN) performed in this work. The number of neurons in the input layer is 9. The 3 hidden layers are used, and the number of neurons in the corresponding hidden layer 1 and 3 are 28 and 71, respectively. The hidden layer 2, namely the dropout layer, has 300 neurons. The number of neutrons in the output layer is 1.} \label{fig1}
\end{figure}
As shown in Fig.~\ref{fig1}, the sketch of Monte Carlo dropout Bayesian neural network has been depicted.
For the input structure, these quantities including the basic building blocks of proton ($Z$) and neutron $(N)$ numbers, isospin asymmetry degree ($I^{2}$), pairing effect $(\delta)$, and quadrupole deformation parameters ($\beta_{20}$), are incorporated into the validated protocol.
Meanwhile, the profound shape staggering phenomena in $^{181,183,185}$Hg ($LI$) isotopes are also labeled in the input layer.
Therefore, the corresponding terms shown in the MCD-BNN model have the following expressions,
\begin{eqnarray}\label{bnn4}
I^{2} &=&(\frac{N-Z}{A})^{2},\\
\delta &=& \frac{(-1)^{Z}+(-1)^{N}}{2},\\
LI &=& \left\{
\begin{array}{c c}
1,&(Z,A)=(80,181),(80,183),(80,185)\\
0,&\mathrm{else}
\end{array}. \right.
\end{eqnarray}
The ground state properties about the quadrupole deformation parameters $\beta_{20}$ are taken from the finite-range droplet model (FRDM)~\cite{MOLLER20161}, relativistic mean field (RMF) theory~\cite{geng2003}, and Weizs\"{a}cker-Skyrme (WS) model~\cite{WANG2014215}.
Meanwhile, the corresponding hyperparameter shown in Eq.~(\ref{bnn3}) is $\lambda_{\mathrm{decay}}=1.6\times10^{-12}$ for the FRDM input. The RMF and WS inputs give the corresponding hyperparameters $\lambda_{\mathrm{decay}}=5.2\times10^{-9}$ and $\lambda_{\mathrm{decay}}=7.3\times10^{-10}$, respectively.

Generally, shell quenching phenomena can be described well by the Casten factor ($P$) which associates with valence neutrons and valence protons~\cite{PhysRevLett.58.658,Angeli_1991,Casten_1996}.
For the calculation of Casten factor $P$ , the reference neutron and proton magic numbers are taken as $Z=2$, $6$, $14$, $28$, $50$, $82$, ($114$) and $N=2$, $8$, $14$, $28$, $50$, $82$, $126$, ($184$).
Actually, the abruptly increasing change can be found in charge radii of Sc isotopes across $N=20$~\cite{PhysRevLett.131.102501}.
In this work, systematic evolution of charge radii of nuclei with proton number $Z\geq20$ and mass number $A\geq40$ is taken into account. The cutoff range of the dataset is in consistent with Refs.~\cite{PhysRevC.105.014308,DONG2023137726,XIAN2025139662}.
Therefore, the reference neutron and proton magic numbers are chosen from the begin of $N(Z)=20$.

As shown in Refs.~\cite{ANGELI201369,LI2021101440}, the shrinking trend of charge radii can be obviously observed around neutron magic numbers along a long isotopic family. Besides, the same scenario can also be encountered at the proton magic numbers along the isotonic chains.
This leads to the inverted parabolic-like shapes of charge radii between the two adjacent magic numbers.
In order to capture these phenomena along an isotopic family, the correlations between the paired valence particles and valence holes have been taken into account in the input structure. Here, the time-reversal symmetry is considered, namely a paired nucleons is became with time-reversal states.
Meanwhile, an assumption of the shell frozen effect is performed, in which the valence particles are not allowed to scatter into higher levels across the next traditional magic numbers.
Such as $^{44}$Ca, the paired valence neutron number is 2, and the corresponding valence hole number is also 2.
This leads to the coupling term between the paired valence particles and valence holes is taken as $2\times2$, which is characterized as $\nu_{\mathrm{n}}^{\mathrm{ph}}$. The self-coupling assumption is performed for the unpaired valence particle, in which the time-reversal symmetry is broken and the self-coupling term for the unpaired particle is taken as fixed value 1. Such as $^{43}$Ca, the coupling term between the paired valence neutrons and valence holes is taken as $1\times2+1$.
The same protocol is also preformed for the valence proton case, the coupling term is marked as $\nu_{\mathrm{p}}^{\mathrm{ph}}$.

As demonstrated in Ref.~\cite{XIAN2025139662}, the traditional Casten factor $P$ is invalid along the isotopic chains or the isotonic chains with the traditional proton or neutron magic numbers.
Therefore, the modified Casten factor $P^{*}$ has been proposed to describe the discontinuous behavior of nuclear charge radii.
In this work, the modified Casten factor $P^{*}$ has been recalled as follows,
\begin{eqnarray}\label{bnn5}
P^{*}&=& \ln\left(\frac{e^{\nu_{\mathrm{p}}^{\mathrm{ph}}}e^{\nu_{\mathrm{n}}^{\mathrm{ph}}}}{{\nu_{\mathrm{p}}^{\mathrm{ph}}}
+{\nu_{\mathrm{n}}^{\mathrm{ph}}}}\right).
\end{eqnarray}
Meanwhile, as mentioned above, the coupling terms between the paired valence particles and valence holes for neutrons and protons are incorporated into this expression.

The final input structure for the MCD-BNN approach is $z=\{Z,$ $N$, $I^{2}$, $\delta$, $P^{*}$, $\nu_{\mathrm{n}}^{\mathrm{ph}}$, $\nu_{\mathrm{p}}^{\mathrm{ph}}$, $\beta_{20}$, $LI\}$.
Meanwhile, the Casten factor is vanished totally if the proton numbers and neutron numbers simultaneously occupy the fully filled shells, such as $(Z=20, N=20)$, $(Z=20, N=28)$, $(Z=28, N=28)$, $(Z=28, N=50)$, $(Z=50, N=50)$, $(Z=50, N=82)$, and $(Z=82, N=126)$ cases.
%Furthermore, the empirical formula shown below~\cite{Nerlo-Pomorska:1994dhg} is chosen as our theoretical model to be refined by MCD-BNN:
%\begin{eqnarray}\label{bnn6}
%R_{\mathrm{np}}(Z,A)=r_{0}A^{1/3}\left(1-b\frac{N-Z}{A}+\frac{c}{A}\right),
%\end{eqnarray}
%where $r_{0}=0.966$ fm, $b=0.182$, and $c=1.652$~\cite{Bayram:2013jua}.

\section{Results and discussion}\label{third}
To depict the local variations of nuclear charge radii, the input-driven analysis is performed by the MCD-BNN model.
In the training set, the $814$ experimental data are taken from Ref.~\cite{ANGELI201369}. The more recent experimental data~\cite{LI2021101440,PhysRevLett.126.032502}, containing $118$ data for nuclei with $Z\geq20$ and $A\geq40$, are used to test the predictive ability of the constructed model in the validation set. Here, the whole data of charge radii along nickel isotopic chain are included into the validation data set.
Meanwhile, charge radii of Bi isotopes shown in Refs.~\cite{PhysRevC.94.024334,PhysRevC.95.044324,sltw-g4d6} are used to examine the extrapolated reliability of MCD-BNN model under various quadrupole deformation contents. This means that these charge radii data cannot be included in the training or validation sets.

The root-mean-square deviation (RMSD) between the results obtained by the MCD-BNN model and the corresponding experimental data is used to quantify the predictive ability, which is shown as follows,
\begin{eqnarray}\label{bnn7}
\sigma^{(\mathrm{T}, \mathrm{V}, \mathrm{TV})}=\sqrt{\frac{1}{N_{(\mathrm{T}, \mathrm{V}, \mathrm{TV})}}\sum_{i=1}^{N_{(\mathrm{T}, \mathrm{V}, \mathrm{TV})}}\left(R_{\mathrm{ch},i}^{\mathrm{Theo.}}-R_{\mathrm{ch},i}^{\mathrm{Expt}.}\right)^{2}}.\nonumber\\
\end{eqnarray}
where $N_{\mathrm{T}}$, $N_{\mathrm{V}}$, and $N_{\mathrm{TV}}$ are the numbers of charge radii data contained in the training (T), validation (V), and entire (TV) data sets, and the subscript $i$ represents the $i$th nucleus in the given data sets.

\begin{figure}[htbp]
\includegraphics[scale=0.35]{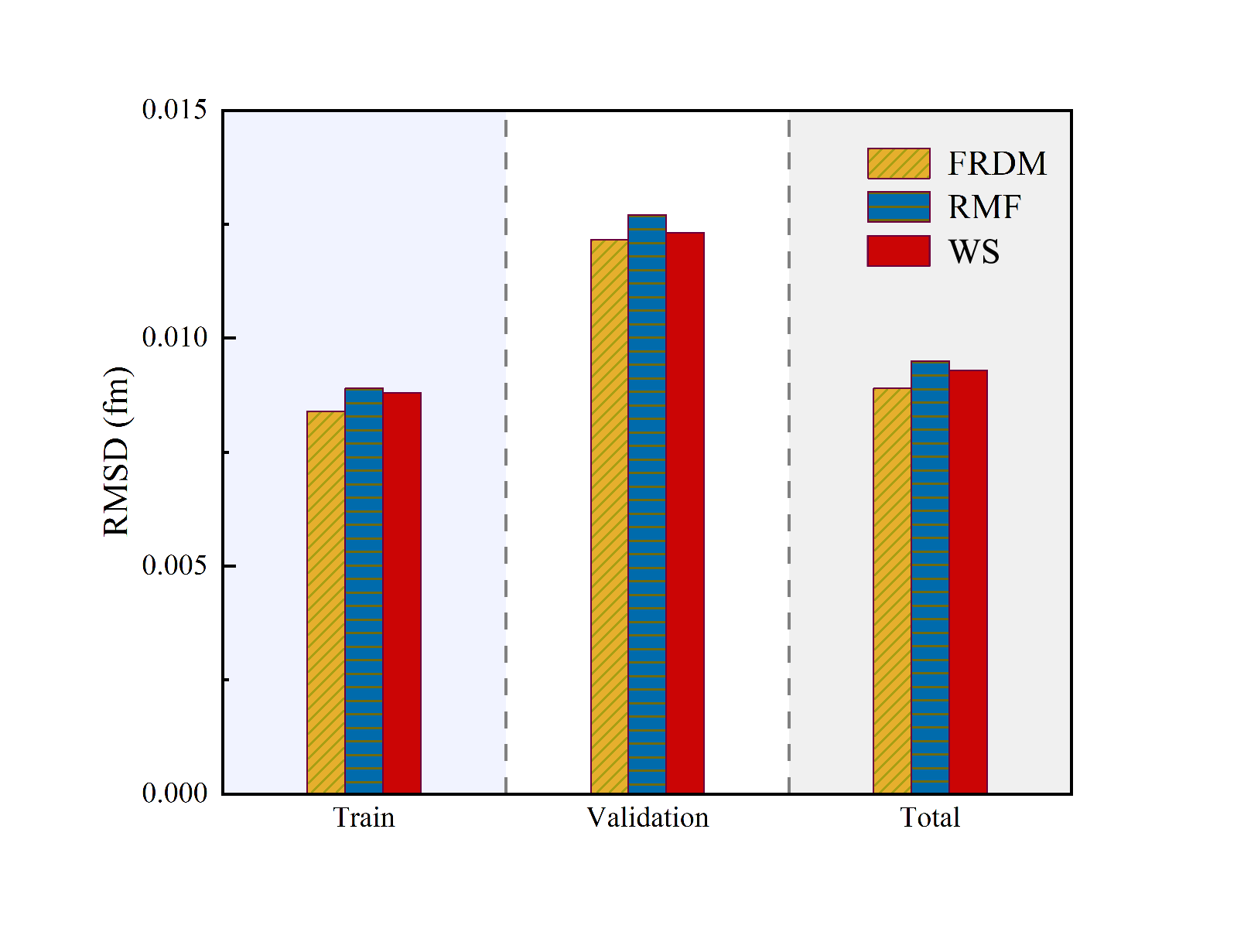}
\caption{Variations of root-mean-square deviations (RMSDs) for the training, validation, and total sets under the framework of FRDM~\cite{MOLLER20161}, RMF~\cite{geng2003} and WS~\cite{WANG2014215}.} \label{fig2}
\end{figure}
To show the quantitative analysis, the influence of the input structure on determining the predictive power is discussed as follows.
The results obtained with various shape deformation contents are used to make a comparative investigation.
As shown in Fig.~\ref{fig2}, the variations of RMSDs for the training, validation, and total sets with various $\beta_{20}$ parameters taken from the finite-range droplet model (FRDM)~\cite{MOLLER20161}, relativistic mean field (RMF) theory~\cite{geng2003} and Weizs\"{a}cker-Skyrme (WS) model~\cite{WANG2014215} are depicted.
As shown in this figure, the FRDM input yields the smallest RMSDs for the training, validation, and total sets compared to the RMF and WS inputs. Specifically, the RMF and WS inputs produce comparable RMSDs in the training and total sets, while the RMF input gives a slightly larger RMSD in the validation set. Table~\ref{tab1} presents the RMSDs of charge radii predicted by the MCD-BNN model using various shape deformation parameters.
From this figure, it can be found that the deviations in the validation sets are relatively large with respect to the training sets for these three models. This can be understood easily that the data including in the validation set come from the nuclei far from $\beta$-stability line. This leads to the relatively large deviation in the validation set due to the absence of more underlying mechanisms.
%The model proves accurate in reproducing the charge radii of known nuclei and reliable in extrapolating to neutron-rich regions that are experimentally uncharted.
%From this figure, one can see that the RMSDs with FRDM input for the train, validation, and total sets are the smallest compared to RMF and WS inputs. Especially, in the train and total sets, RMF and WS inputs give comparable RMSDs.
%In the validation set, a slightly larger RMSD can be obtained by the RMF input.
%As shown in Table~\ref{tab1}, the RMSDs of charge radii predicted by the MCD-BNN model with various shape deformation parameters are given.
%This model is demonstrated to be accurate in reproducing the charge radii of experimentally known nuclei and reliable in extrapolating to the experimentally unknown neutron-rich regions.

\begin{table}[htb!]
\centering
\caption{Root-mean-square deviations (RMSDs) of charge radii predicted by the MCD-BNN model with various shape deformation parameters  taken from FRDM~\cite{MOLLER20161}, RMF~\cite{geng2003} and WS~\cite{WANG2014215} models, respectively.  }\label{tab1}
\doublerulesep 0.1pt \tabcolsep 6.0 pt
\begin{tabular}{cccccc}
\hline
\hline
Model & $\sigma^{(\mathrm{T})}$~(fm)  & $\sigma^{(\mathrm{V})}$~(fm) & $\sigma^{(\mathrm{TV})}$~(fm) \\
\hline
MCD-BNN(FRDM) &0.0084  & 0.0122  & 0.0089  \\
MCD-BNN(RMF)  &0.0089 & 0.0127  & 0.0095   \\
MCD-BNN(WS)   &0.0088  & 0.0123 & 0.0093   \\
\hline
\hline
\end{tabular}
\end{table}

\begin{figure}[htbp]
\includegraphics[scale=0.8]{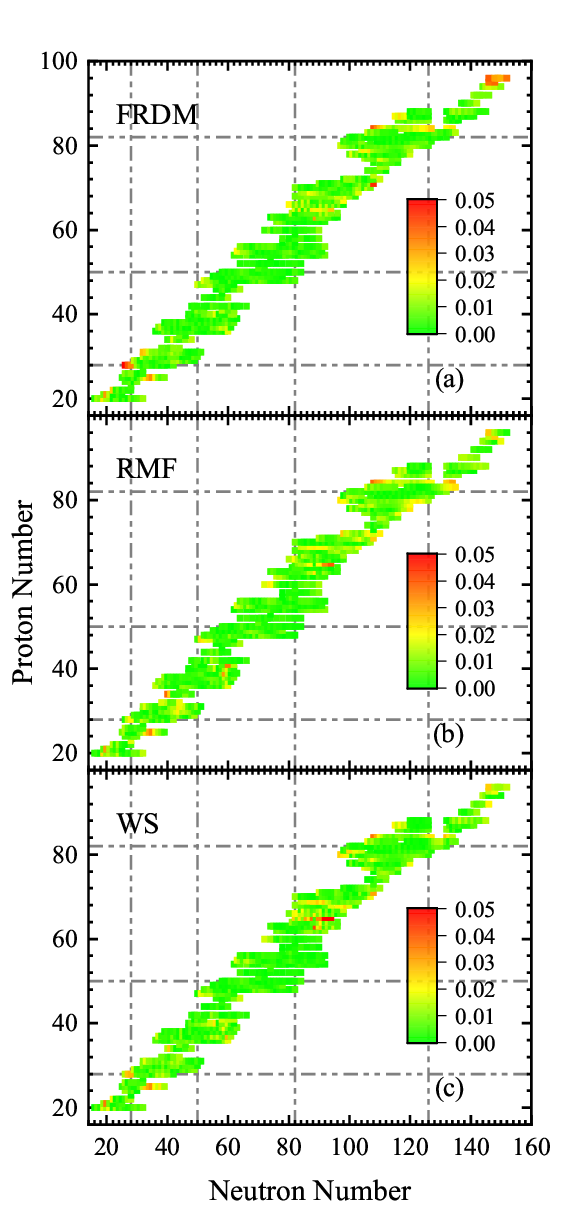}
\caption{Nuclear chart displaying the absolute differences (in units of fm) between the rms charge radii obtained by the MCD-BNN model with the input-driven deformation parameters taken from FRDM~\cite{MOLLER20161}, RMF~\cite{geng2003} and WS~\cite{WANG2014215} models and the corresponding experimental data in the entire set. The results are obtained from the training
and validation sets, respectively. } \label{fig3}
\end{figure}

To depict the training and validation sets of MCD-BNN model in details, as shown in Fig.~\ref{fig3}, the difference between the calibrated rms charge radii with the input-driven deformation parameters taken from FRDM~\cite{MOLLER20161}, RMF~\cite{geng2003}, and WS~\cite{WANG2014215} models and the corresponding experimental data are depicted in nuclear chart.
It should be obviously mentioned that most of nuclear charge radii data can be reproduced well.
However, in some specific regions, the calibrated results deviate from the experimental data.
Particularly, around $Z=65$ isotopic chains, more larger deviations can be obtained by various shape deformation inputs.
This can be easily understood that the inverse odd-even staggering phenomena are significantly observed in these regions~\cite{AHMAD1988244,Alkhazov337_257,PhysRevC.99.054317,PhysRevC.100.044321}, but this case cannot be incorporated into input layer.
In addition, larger deviations can also be encountered around $Z=82$ isotopic chains toward neutron-deficient regions, although the shape staggering phenomena have been incorporated into the input structure.
Meanwhile, it should be significantly mentioned that larger deviations can also be obtained close to neutron-rich side in heavy regions.
As is well known, high-order octupole deformation effect has been predicted in the northeast regions of nuclide chart and has an influence on determining the bulk properties of heavy nuclei~\cite{RevModPhys.68.349,MOLLER2008758,Gaffney2013,Butler_2016,PhysRevC.93.044304,PhysRevC.102.024311,Butler2020}.
In our calculations, the quadrupole deformation has been incorporated into the input structure.
This leads to more larger deviations of RMSD in calibrating nuclear charge radii due to the absence of high-order octupole deformation.

\begin{figure}[htbp]
\centering
\includegraphics[scale=0.5]{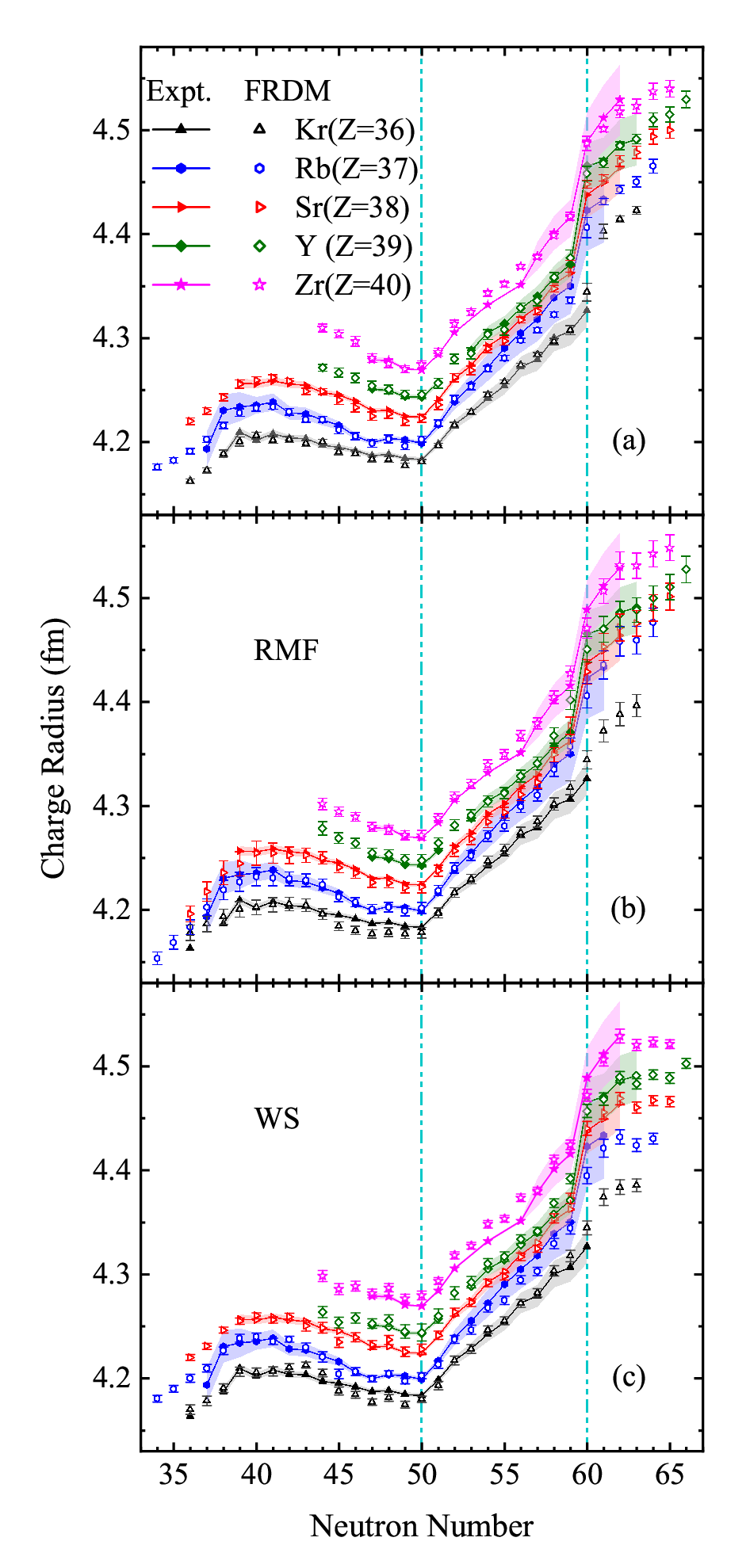}
\caption{Charge radii of Kr, Rb, Sr, Y, and Zr isotopic chains are depicted by the MCD-BNN model using the FRDM (a)~\cite{MOLLER20161}, RMF (b)~\cite{geng2003} and WS (c)~\cite{WANG2014215} inputs. The experimental data are taken from Refs.~\cite{ANGELI201369,LI2021101440}.} \label{fig4}
\end{figure}
It is also important to feature the reliability of MCD-BNN model when extrapolated to the discontinuous regions of charge radii with abrupt shape phase transition.
To characterize the local variations of charge radii, the systematic evolution of charge radii along Kr, Rb, Sr, Y, and Zr isotopic chains are depicted in Fig.~\ref{fig4}.
As shown in this figure, shell quenching phenomena at $N=50$ can be reproduced well using various input-driven deformation structures.
In contrast to the FRDM and RMF inputs, charge radii of $^{92-96}$Zr isotopes simulated with WS input slightly deviate from the experimental data.
Around $N=60$ region, the abrupt increase of charge radii can be reproduced well by the MCD-BNN models with FRDM, RMF, and WS inputs.
Along Kr isotopes, the profoundly increasing trend of charge radii cannot be significantly predicted in these three inputs compared to Rb, Sr, Y, and Zr isotopic chains.
This is consistent with Ref.~\cite{PhysRevC.108.024310} where the abrupt increase of charge radii of Kr isotopes cannot be predicted by the deformed relativistic Hartree-Bogoliubov theory in continuum at $N=60$.
Here, it should be mentioned that the calibrated charge radii of $^{96}$Kr isotope using FRDM, RMF, and WS inputs are slightly overestimated, although the simulated results cover partly the uncertainty range of experimental datum.
Besides, the FRDM input predicts the rapid trend of charge radii from $^{96}$Kr to $^{97}$Kr compared to RMF and WS inputs.
Thus more charge radii data should be urgently required in this region.
Close to proton-rich regions, the RMF input overestimates the charge radii of Kr isotopes with respect to the FRDM and WS inputs.
Particularly, charge radius of $^{72}$Kr isotope is obviously overestimated by the RMF input.

Actually, the rapidly increasing trend of changes of charge radii is weakened along $Z=41$ isotopic chain as well in our simulations.
For the FRDM input, charge radii of $^{100}$Nb and $^{101}$Nb isotopes are predicted as 4.4361(25) fm and 4.4798(67) fm, respectively.
With considering the RMF input, charge radii of $^{100}$Nb and $^{101}$Nb isotopes are 4.4286(123) fm and 4.4507(127) fm, respectively.
Meanwhile, the WS input gives the values of charge radii of $^{100}$Nb and $^{101}$Nb isotopes 4.4366(60) fm and 4.4714(66)fm, respectively.
In contrast to the Rb, Sr, Y, and Zr isotopic chains, this increasing trend of changes of charge radii around $N=60$ is remarkably reduced in Nb isotopes.
From these simulations, it seems to demonstrate that the abrupt shape-phase transition regions around $N=60$ can be determined from the rapid trend of nuclei size.

As shown in Fig.~\ref{fig4}, three inputs give almost similar evolution curves of charge radii along Kr, Rb, Sr, Y, and Zr isotopic chains until to neutron number $N=60$.
Beyond $N=60$, the similar increasing trend can be predicted by the FRDM and RMF inputs along Kr, Rb, Sr, Y, and Zr isotopes.
However, for the WS input, this increasing trend is weakened or the flat trend is given except for Kr isotopes.
It should be mentioned that quadrupole deformation parameters $\beta_{20}$ incorporated into the MCD-BNN framework are different for the FRDM, RMF, and WS models.
%Thus the slightly deviations can be predicted in theoretical simulations.
%Therefore, the predicted trend of changes of charge radii in neutron-rich regions is different among three inputs.
%This means that shape deformation plays an indispensable role in determining the systematic evolution of nuclear charge radii.
Therefore, the slight deviations predicted in theoretical simulations lead to differing trends in the charge radii for neutron-rich regions among these three inputs. This indicates that shape deformation plays an indispensable role in determining the systematic evolution of nuclear charge radii.

The systematic evolution of charge radii along Bi isotopic chain attracts more attention in recent investigation.
The charge radii of the $^{193,195,197}$Bi isotopes have been detected~\cite{PhysRevC.94.024334}.
In this study, the larger charge radii can be obtained for the isomers states due to the larger shape deformation. Furthermore, the shape coexistence phenomena should be considered properly in describing the systematic behavior of nuclear charge radii.
%Particularly, toward neutron-deficient regions, the quadrupole deformation plays an important role in featuring the larger odd-even staggering phenomena of charge radii along Bi isotopic chain~\cite{PhysRevC.95.044324}.
%Recently, charge radii of $^{214\ensuremath{-}218}\mathrm{Bi}$ isotopes have been measured using the in-source resonance-ionization spectroscopy technique at ISOLDE (CERN)~\cite{sltw-g4d6}. The observed results demonstrate the normal odd-even staggering in charge radii of Bi isotopes beyond $N=126$ as well. Meanwhile, it should be remarkably mentioned that high-order octupole deformation can have an influence on determining the locally discontinuous variations of charge radii.
Particularly, in neutron-deficient regions, the quadrupole deformation plays an important role in causing the pronounced odd-even staggering of charge radii along the Bi isotopic chain~\cite{PhysRevC.95.044324}. Recently, charge radii of $^{214\ensuremath{-}218}\mathrm{Bi}$ isotopes have been measured using the in-source resonance-ionization spectroscopy technique at ISOLDE (CERN)~\cite{sltw-g4d6}. The results demonstrate that a normal odd-even staggering also exists in the charge radii of Bi isotopes beyond $N=126$. Meanwhile, it is important to note that high-order octupole deformation can also influence locally discontinuous variations of charge radii.

\begin{figure}[htbp]
\centering
\includegraphics[scale=0.24]{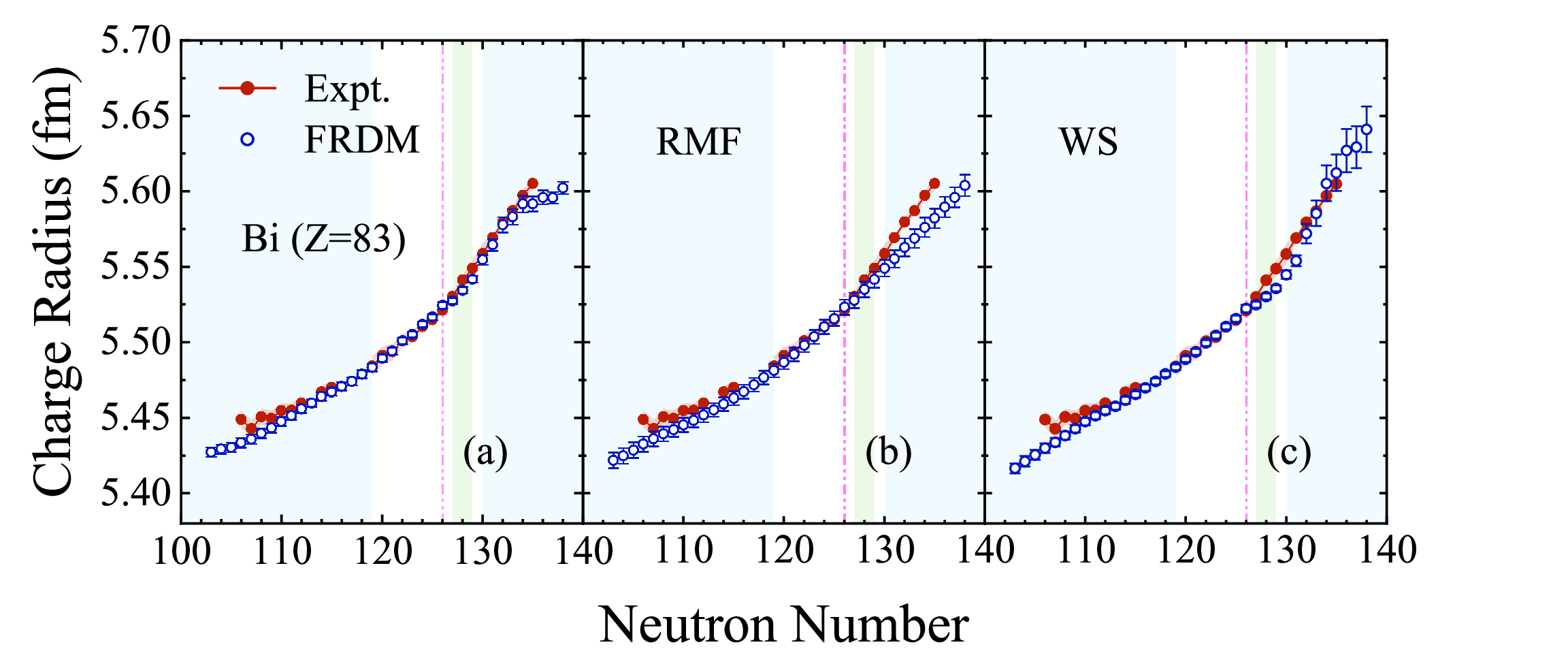}
\caption{Charge radii of Bi isotopes are depicted by the MCD-BNN model using the FRDM (a)~\cite{MOLLER20161}, RMF (b)~\cite{geng2003} and WS (c)~\cite{WANG2014215} inputs. The experimental data are taken from Refs.~\cite{ANGELI201369,PhysRevC.94.024334,PhysRevC.95.044324,LI2021101440,sltw-g4d6}. The regions marked by light blue represents test sets. The training set remark without color and light green labels validation set.} \label{fig5}
\end{figure}

To further review the local variations of charge radii, as shown in Fig.~\ref{fig5}, charge radii of Bi isotopes are depicted by the MCD-BNN model with FRDM, RMF, and WS input structures, respectively.
From this figure, one can see that charge radii of $^{201-211}$Bi isotopes can be reproduced well by the FRDM input, although the slight deviations can be predicted at the neutron numbers $N=131-134$.
The same scenarios can also be obtained by the WS input, namely the slight deviations can be found around $N=130$.
The RMF input can reproduce the charge radii of $^{201-213}$Bi isotopes well.
However, with the increasing neutron numbers, the deviations between the experimental data and theoretical simulations are gradually increased compared to the FRDM and WS inputs.
This can be easily understood that the quadrupole deformation parameters predicted by the FRDM, RMF, and WS models are different.
Such as for $^{218}$Bi isotope, FRDM and WS models give the quadrupole deformation parameters are $0.056$ and $0.020$, respectively, but the RMF model gives the value of $-0.025$.
It should be mentioned that the FRDM, RMF, and WS input structures can reproduce the shell closure effect of charge radii at the fully filled shell $N=126$.
As the aforementioned discussion, the high-order octupole deformation can be generally predicted at the $130\leq{N}\leq140$
regions~\cite{RevModPhys.68.349,MOLLER2008758,Gaffney2013,Butler_2016,PhysRevC.93.044304,PhysRevC.102.024311,Butler2020}.
In our calculations, only quadrupole deformation contents have been incorporated into the input layer.
This seems to result in the slight deviations of charge radii between the experimental data and theoretical simulations around $N=130$ regions.
Therefore, a hint of the reliability of the MCD-BNN model is that high-order shape deformation should be considered in calibrating the charge radii of nuclei far from $\beta$-stability line.

Toward neutron-deficient regions, the theoretical simulations gradually deviate from the experimental data under various input structures.
Especially, the remarkable odd-even staggering phenomena cannot be reproduced well in this region, although the shape staggering phenomena have been marked in the calibrated protocol.
This further suggests that more reliable shape deformation inputs should be considered in the neural network validation.
From the aforementioned discussion, one can find that the root-mean-square deviations between the experimental data and theoretical simulations are further reduced with respect to the existing literature.
This manifests the reliability of the MCD-BNN model in refining nuclear charge radii prediction.
Table.~\ref{tab2} summarizes the extrapolated charge radii of $^{196,199\ensuremath{-}201}\mathrm{Bi}$ isotopes as follows.
\begin{table}[htb!]
\centering
\caption{Charge radii of $^{196,199\ensuremath{-}201}\mathrm{Bi}$ isotopes (in units of fm) predicted by the MCD-BNN model are shown with the associated uncertainty. The various shape deformation parameters incorporated into the input structure are taken from FRDM~\cite{MOLLER20161}, RMF~\cite{geng2003} and WS~\cite{WANG2014215} models, respectively.  }\label{tab2}
\doublerulesep 0.1pt \tabcolsep 10.8 pt
\begin{tabular}{cccccc}
\hline
\hline
Nuclei & FRDM~(fm)  & RMF~(fm) & WS~(fm) \\
\hline
 $^{196}\mathrm{Bi}$  &5.4598(27) & 5.4552(46)  & 5.4577(23)  \\
 $^{199}\mathrm{Bi}$  &5.4705(26) & 5.4672(45)  & 5.4699(21)   \\
 $^{200}\mathrm{Bi}$  &5.4738(27)  & 5.4716(45) & 5.4742(21)   \\
 $^{201}\mathrm{Bi}$  &5.4786(26)  & 5.4763(45) & 5.4790(19)   \\
\hline
\hline
\end{tabular}
\end{table}
In addition, the latest charge radii data along Sn isotopic chain are also used to measure the extrapolation capability of these constructed models~\cite{PhysRevLett.135.222501}. The calculated results suggest that systematic evolution of charge radii along Sn isotopic chain can also be described well, but slight deviations can be encountered for the FRDM inputs close to neutron-deficient regions. This can be understood easily that the quadrupole deformation parameters in these regions are different for three input contents.

In this work, it should be mentioned that larger deviations appear in some specific regions due to the varying quadrupole deformation inputs.
Thus more available description of quadrupole moments plays an essential role in describing the bulk properties of finite nuclei~\cite{frfs-xz86}.
Besides, high-order shape deformation contents and inverse odd-even staggering phenomena cannot be incorporated into the input structure.
These effects can also have an influence on the determination of nuclear charge radii.
It is known that the charge radii for most neutron-rich nuclei are still beyond the experimental capability.
One may improve the predictions of nuclear charge radii with the help of machine-learning approaches when more motivated underlying mechanisms are considered properly in the input structure.
\section{Summary and outlook}\label{fourth}

In this work, input-driven Monte Carlo dropout Bayesian neural networks have been built to describe the systematic evolution of charge radii of nuclei with proton number $Z\geq20$ and mass number $A\geq40$.
In input layer, the motivated underlying mechanisms, which include proton numbers, neutron numbers, pairing effect, isospin asymmetry degree, correlations between the valence particles and valence holes for proton and neutron, shell closure effect, the marked shape staggering of $^{181,183,185}$Hg, and quadrupole deformation, are considered in the calibrated process.
The shape deformation contents derived from the finite-range drop model~\cite{MOLLER20161}, relativistic mean field theory~\cite{geng2003}, and Weizs\"{a}cker-Skyrme approach~\cite{WANG2014215} have been added into input layer.
Comparison of our results for charge radii with different shape deformation inputs is made. The calculated results suggest that the comparable root-mean-square deviations are obtained in the training and validation sets.
This model is demonstrated to be available in reproducing the charge radii of experimentally known nuclei and reliable in extrapolating to the experimentally unknown neutron-rich regions.
However, it should be mentioned that more larger deviations can be encountered due to the absence of inverse odd-even staggering mechanism and available shape deformation contents.

Shell quenching phenomena of charge radii can be reproduced well at $N=50$ along Kr, Rb, Sr, Y, and Zr isotopic chains.
The abrupt increasing trend of changes of charge radii can be reproduced well along $Z=37-40$ isotopic chains.
In contrast, this rapid increasing trend is weakened along $Z=36$ and $41$ isotopic chains.
This seems to clarify the shape-phase transition regions around $N=60$ from nuclei size perspective.
Furthermore, the evolution curves of charge radii show a similar behavior along Kr, Rb, Sr, Y, and Zr isotopes until to $N=60$.
Beyond $N=60$, evolution curves of charge radii obtained by WS input are different from the cases obtained by the FRDM and RMF inputs due to the distinguished quadrupole shape deformation contents.
This means that a proper input structure plays an important role in calibrating the theoretical simulations.

The predictive charge radii of Bi isotopes slightly deviate from the experimental data around $N=130$ and toward neutron-deficient regions, although the shell closure effect can be described well at the fully filled $N=126$ shell.
As the aforementioned discussion, the larger shape staggering information cannot be adequately captured in the calibrated process, although the corresponding information is marked along $^{181,183,185}$Hg isotopes.
Besides, high-order octupole deformation should be considered in predicting the nuclear charge radii.
As is well known, structure phenomena can be remarkably presented close to light regions of nuclide chart. In our calculations, the simulated charge radii data are chosen for nuclei with $Z\ge20$ and $A\ge40$. Thus more relatively light nuclei should be taken into account in the proceeding research.
This means that more motivated underlying mechanisms should be incorporated into the input structure properly.

\section{Acknowledgements}
This work was supported by the key research and development project of Ningxia, China (Grants No. 2024BEH04090), the Natural Science Foundation of Ningxia Province, China (Nos. 2024AAC03015 and 2025AAC030208), the Central Government Guidance Funds for Local Scientific and Technological Development, China (No. Guike ZY22096024), the Open Project of Guangxi Key Laboratory of Nuclear Physics and Nuclear Technology (No. NLK2023-05), and the Key Laboratory of Beam Technology of Ministry of Education, China (Nos. BEAM2024G04 and BEAM2024G05).

%\bibliography{refsanw}

\end{document}